\documentclass[12pt]{article}
\usepackage{url}
\input{epsf}

\font\bm=cmmib10 at 10pt
\font\bms=cmmib10 at 7pt \textfont9=\bm \scriptfont9=\bms
\usepackage{amsfonts}
\usepackage{multirow}
\mathchardef\balpha= "790B
\mathchardef\bbeta= "790C
\mathchardef\bTheta= "7902
\mathchardef\bzeta= "7910
\mathchardef\bOmega= "790A
\mathchardef\bGamma= "7900
\mathchardef\bDelta= "7901
\mathchardef\bPhi= "7908
\mathchardef\bphi= "791E
\mathchardef\bomega= "7921
\mathchardef\bxi= "7918
\mathchardef\bet= "7911
\mathchardef\brho= "791A
\mathchardef\btau= "791C
\mathchardef\bmu= "7916
\mathchardef\bvarpi= "7924

\def \lvec{(\kern-.26em(}
\def\pmb#1{\setbox0=\hbox{#1}%
\def \lvec{(\kern-.26em(}
\kern-.025em\copy0\kern-\wd0
\kern.05em\copy0\kern-\wd0
\kern-.025em\raise.0433em\box0 }
\mathchardef\btheta= "7912
\usepackage{amsmath}
\usepackage{authblk}

\usepackage{graphicx}
\usepackage{dcolumn}
\usepackage{rotating}

\begin{document}

\title{Instantaneous Clear Sky Radiative Forcings of Halogenated Gases}
\author[1]{W. A. van Wijngaarden}
\author[2]{W. Happer}
\affil[1]{Department of Physics and Astronomy, York University, Canada, wavw@yorku.ca}
\affil[2]{Department of Physics, Princeton University, USA, happer@Princeton.edu}
\renewcommand\Affilfont{\itshape\small}
\date{\today}
\maketitle

\noindent  The clear sky instantaneous radiative forcings of the 14 halogenated gases previously shown to have the largest contribution to global warming, were found.  The calculation used the absorption cross sections for the halogenated gases which are assumed to be independent of temperature as well as over 1/3 million line strengths for the five main naturally occurring greenhouse gases: H$_2$O, CO$_2$, O$_3$, CH$_4$ and N$_2$O, from the Hitran database.  The total radiative forcing of the halogenated gases at their 2020 concentrations is 0.52 (0.67) W/m$^2$ at the tropopause (mesopause).  Over half of this forcing is due to CFC11 and CFC12 whose concentrations are steadily declining as a result of the Montreal Protocol.  The rate of total forcing change for all 14 halogenated gases is 1.5 (2.2) mW/m$^2$/year at the tropopause (mesopause).  The calculations assumed a constant altitude concentration for all halogenated gases except CFC11, CFC12 and SF$_6$.  Using the observed altitude dependence for those 3 molecules reduced their radiative forcings by about 10\%.  The global warming potential (GWP) values were found to be comparable to those given by the Intergovernmental Panel on Climate Change.  The contribution of a gas to global warming can be estimated using the forcing power per molecule defined as the derivative of its radiative forcing with respect to its column density.  For the present atmosphere, the per-molecule forcing powers of halogenated gases are orders of magnitude larger than those for the 5 naturally occuring greenhouse gases because the latter have much higher concentrations and are strongly saturated.  However, the rates of concentration increase of the 5 main greenhouse gases are orders of magnitude greater than that of any halogenated gas.  Assuming the temperature increase caused by each gas is proportional to its radiative forcing increase, the 14 halogenated gases are responsible for only 2\% of the total global warming.

%
\newpage
\section{Introduction}
Halogenated molecules frequently contain one or more carbon atoms as well as various halogen atoms and hydrogen.  Chlorofluorocarbons (CFCs) are nontoxic and nonflammable.  CFC11 and CFC12, commercially known as freon 11 and freon 12 respectively, were discovered about a century ago and widely used in refrigeration.  When released into the atmosphere, these molecules can travel to the stratosphere where they break down releasing chlorine atoms.  In the 1970s, it was found that chlorine catalyzed the destruction of the stratospheric ozone layer \cite{Molina, Crutzen}.  An international agreement known as the Montreal Protocol was created in 1987 to eliminate the use of CFCs \cite{Montreal}.  Subsequent amendments seek to phase out hydrochlorofluorocarbons (HCFCs) and hydrofluorocarbons (HFCs).  These molecules were developed to replace CFCs.  They have much less ozone destroying ability but along with the CFCs are greenhouse gases as pointed out by the Integovernmental Panel on Climate Change (IPCC) \cite{IPCC2021}.   

There have been a number of calculations of the radiative forcings and global warming potentials for various halogenated molecules \cite{Jain, Sihra, Buehler2022}.  These calculations used various radiative transfer models along with results from the Hitran database for the absorption cross sections for the halogenated gases as well as line strengths for greenhouse gases such as carbon dioxide \cite{HITRAN}.  The calculations use temperature profiles and altitudinal profiles of greenhouse gases such as water vapor at various locations on the Earth.  Unfortunately, these profiles are not always readily available which would facilitate comparison of different calculations.  

The instantaneous radiative forcing was calculated for 14 molecules found to be the strongest greenhouse gases of the halogenated molecules.
The calculations were done for a clear sky and use the standard US temperature profile for the midlatitude \cite{Temp}.  This expands our previous work that found the radiative forcings of the 5 main naturally occurring greenhouse gases H$_2$O, CO$_2$, O$_3$, CH$_4$ and N$_2$O \cite{WvW1}.  Our calculations use the observed altitudinal profiles of these gases \cite{Anderson}.  The radiative forcings of the halogenated gases using their 2020 concentrations, were found at the tropopause and top of the atmosphere or mesopause.  The rate of change of the total halogenated gas forcing was found using the observed concentration change rates of the various gases.  The warming effectiveness of each molecule was found by taking the derivative of its radiative forcing with respect to its column density to obtain the per-molecule forcing power.  Finally, the global warming potentials were obtained and compared to values in the literature.

\section{Concentration of Greenhouse Gases \label{he}}

Careful measurements of surface greenhouse gas concentrations have been ongoing for more than half a century.  The pioneering work of C. Keeling studying carbon dioxide concentrations at Mauna Loa has been extended to include a number of halogenated species as shown in Fig. 1 \cite{MaunaLoa}.  
The graph clearly shows the effectiveness of the Montreal Protocol as the concentrations of the two most abundant halogenated gases, CFC11 and CFC12 have decreased by about 20\% and 10\% respectively, since their peak concentrations in the 1990s.  All of the halogenated molecules have an entirely anthropogenic origin with the exception of CF$_4$ whose preindustrial level is estimated to be 34 ppt \cite{MaunaLoa}.

\begin{figure}[t]
\includegraphics[height=120mm,width=1
\columnwidth]{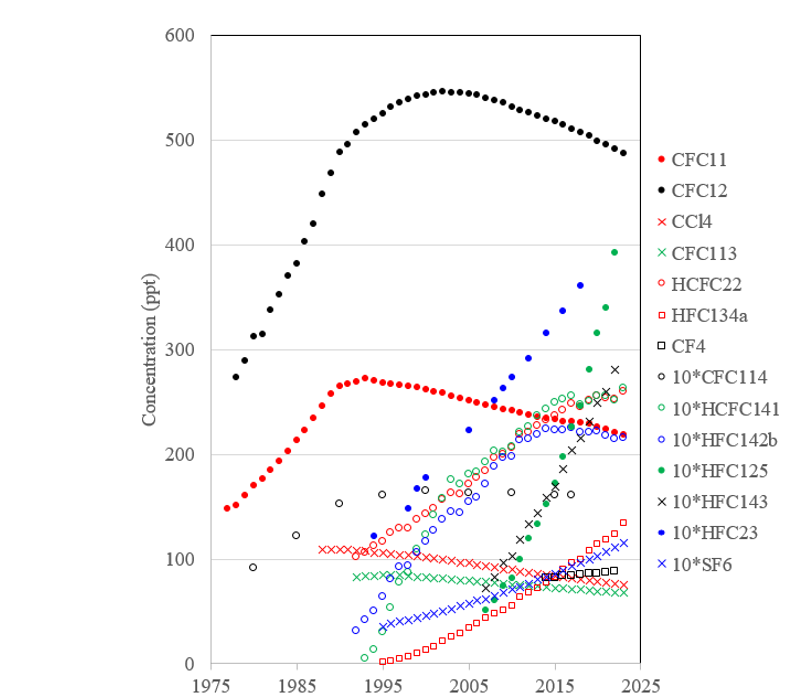}
\caption{Measurements of Halogenated Gas Concentrations.  The data for CFC114 are taken from Vollmer et al \cite{Vollmer} while those for HFC23 are from Stanley et al \cite{Stanley}.  The data for all other gases were obtained by NOAA's Global Greenhouse Gas Reference Network at Mauna Loa, Hawaii \cite{MaunaLoa}.  Note the concentrations of several gases have been multiplied by 10 in order to be visible on the graph as indicated in the legend.
\label{HalogenConcb}}
\end{figure}

\begin{figure}[t]
\includegraphics[height=100mm,width=1\columnwidth]{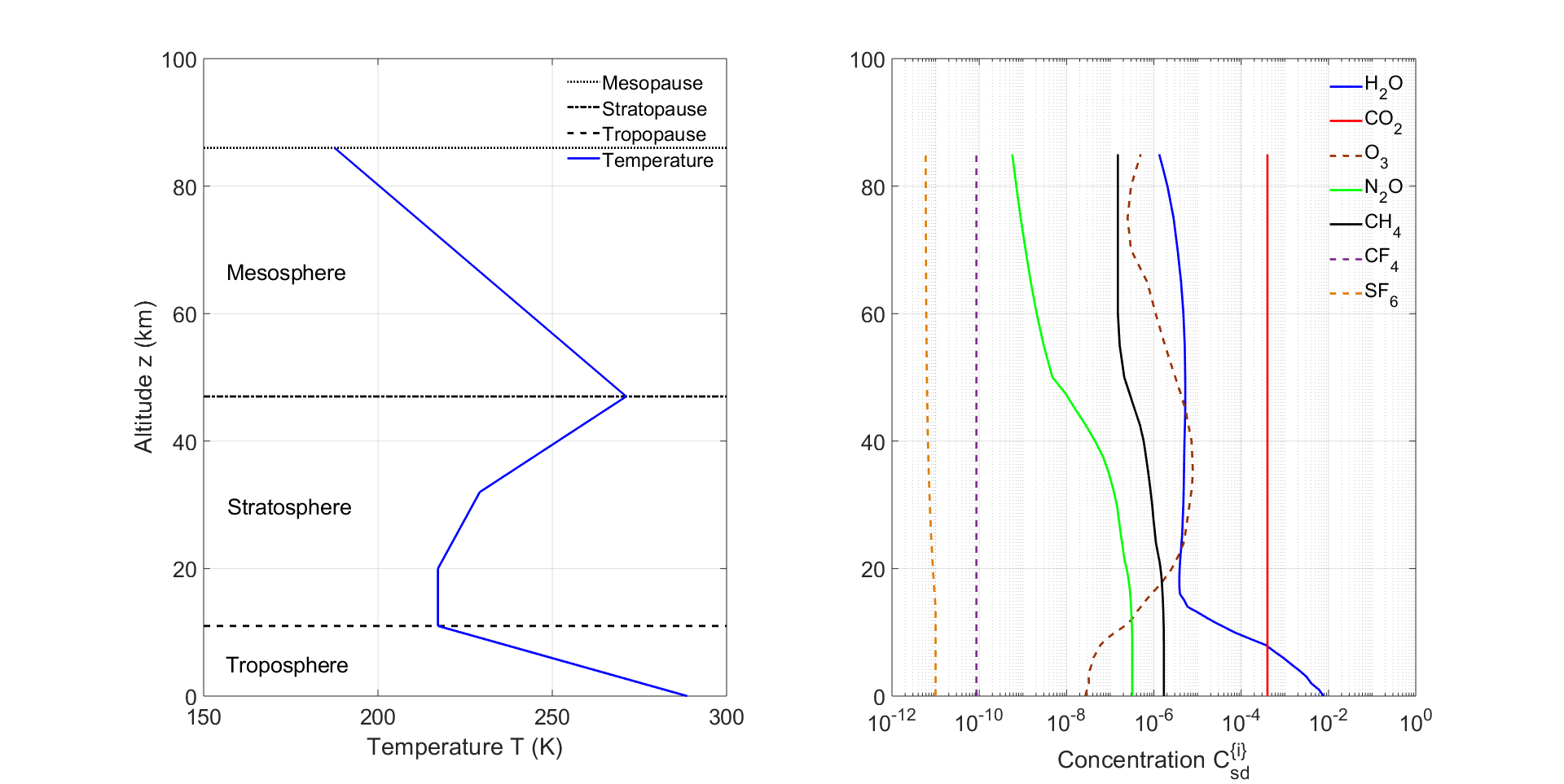}
\caption{{\bf Left.} A midlatitude atmospheric temperature profile, $T=T(z)$. The Earth's mean surface temperature  $T(0) = 288.7$ K.  {\bf Right.} Standard observed concentrations, $C^{\{i\}}_{\rm sd}$ for greenhouse molecules versus altitude $z$. 
\label{GGNT}}
\end{figure}

Radiation transfer in the cloud-free atmosphere of the Earth is controlled by the temperature $T=T(z)$ at the altitude $z$ and the number densities, $N^{\{i\}}=N^{\{i\}}(z)$ of the $i$th type of molecule.  
Representative midlatitude altitude profiles of temperature \cite{Temp}, and greenhouse gas concentrations \cite{Anderson}, are shown in Fig. \ref{GGNT}.  Altitude profiles directly measured by radiosondes in ascending balloons are always more complicated than those of Fig. \ref{GGNT}, which can be thought of as appropriate average profiles for the year 2020.  

The standard US temperature profile for the midlatitude, plotted as the solid blue line in the left panel of Fig. 2, consists of 5 linear segments.  The average surface temperature of the Earth is 288.7 K.  This decreases to 217.2 K at the tropopause altitude, $z_{\rm tp}$, of 11 km and remains constant until 20 km.  The temperature in the stratosphere increases to 229.2 K at 32 km with an additional increase to 271.2 K at 47 km as a result of absorption of incoming solar UV radiation by ozone molecules.  Finally, the temperature decreases in the mesosphere to 187.5 K at the mesopause altitude, $z_{\rm mp}$, of 86 km.

The concentrations for the ith greenhouse gas, $C^{\{i\}}_{\rm sd}$, based on observations \cite{Anderson}, are shown as functions of altitude on the right of Fig. \ref{GGNT} for the 5 main greenhouse gases.  The 2020 surface concentrations are $7,750$ ppm of H$_2$O, $1.9$ ppm of CH$_4$, $0.33$ ppm of N$_2$O and 10 ppt for SF$_6$ \cite{MaunaLoa}. The O$_3$ concentration peaks at $7.8$ ppm at an altitude of 35 km, while the concentrations of CO$_2$ and CF$_4$ were 413 ppm and 86 ppt, respectively, at all altitudes.  

The dependence of the concentrations relative to the surface of CFC11 and CFC12 with altitude is shown in Fig. {\ref{ACECFC1112}.  The concentrations of other halogenated gases undoubtedly also depend on altitude but unfortunately few observations exist.  This study considered a constant altitude dependence of the halogenated gas concentrations other than for CFC11, CFC12 and SF$_6$.    

Integrating the number density, $N^{\{i\}}$, over an atmospheric column having a cross sectional area of 1 cm$^2$ yields the column number density of the $i$th type of molecule $\hat N^{\{i\}}_{\rm sd}$ listed in Table 2.

\begin{figure}[t]
\includegraphics[height=100mm,width=.9\columnwidth]{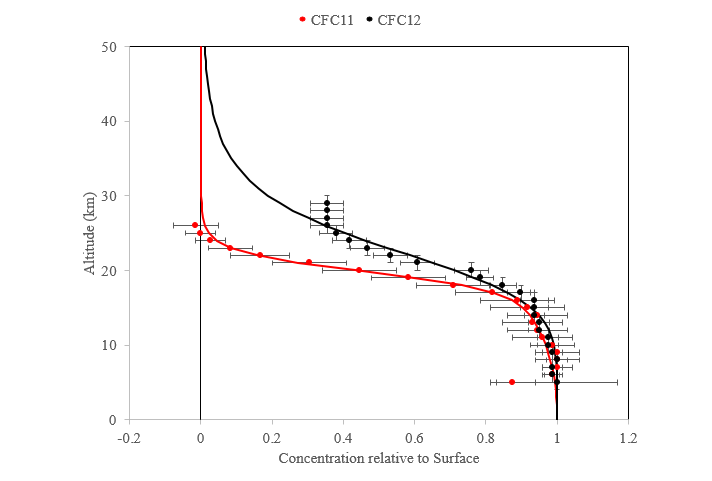}
\caption{Observed concentrations of CFC11 and CFC12 relative to the surface were found using data taken by the ACE Fourier Transform Spectrometer in 2009 between latitudes of 30$^o$ and 45$^o$ N \cite{ACE}. 
\label{ACECFC1112}}
\end{figure}

\section{Calculation of Radiative Forcing}

Our previous publications describe in detail how to calculate radiation transport from the Earth's surface through the atmosphere \cite{WvW1,WvW2} and is here only briefly discussed.  Radiation transport is governed by the Schwarzschild equation \cite{Schwarzschild1906} in cloud-free air where scattering is negligible.

\begin{equation}
	\cos \theta \frac{\partial \tilde I}{\partial \tau}=\tilde B-\tilde I
	\label{nsa4}
\end{equation}

\noindent Here ${\tilde I} = {\tilde I}(\nu,z,\theta)$ is the spectral intensity of a pencil of radiation of frequency between $\nu$ and $\nu + d\nu$ at altitude $z$.  The pencil makes an angle $\theta$ to the vertical.  
The Planck intensity is given by 

\begin{equation}
	\tilde B(\nu,T)=\frac{2h c^2\nu^3}{e^{\nu c\, h/(k_{\rm B}T)}-1}
	\label{b12}
\end{equation}

\noindent The radiation frequency, $\nu$ has units of cm$^{-1}$, $h$ is Planck's constant, $k_B$ is Boltzmann's constant and $c$ is the speed of light.

The optical depth is defined by 

\begin{equation}
	\tau(z,\nu)=\int_0^{z}dz' \kappa(z',\nu),
	\label{b10}
\end{equation}

\noindent where the net attenuation coefficient due to molecules absorbing and reemitting light of frequency $\nu$ at altitude $z$ is given by 

\begin{equation}
	\kappa(z,\nu)=\sum_i N^{\{i\}}(z)\sigma^{\{i\}}(z,\nu).
	\label{b8}
\end{equation}

\begin{table}
\begin{center}
\begin{tabular}{|l|c|c|c|c|c|}
\hline
\ \ \ \ Molecule &Temperature &Pressure & $\nu_{min}$ &$\nu_{max}$ &Cross Section \\ 

&(K) &(mbar)  &(cm$^{-1}$) &(cm$^{-1}$) &($10^{-17}$ cm$^2$) \\
[0.5ex]
\hline\hline
Chlorofluorocarbons& & & & &\\
\ \ CFC10 (CCl$_4$) &295.7 &760.0 &700 &860 &6.73\\
&&&&&\\	
\ \ CFC11    &293.0 &760.0 &710 &1290 &9.95\\
\ \    &273.0 &7.5  &710 &1290 &9.96\\
\ \    &232.6 &7.5  &710 &1290 &9.96\\
\ \    &216.7 &49.9 &710 &1290 &9.96\\
\ \    &216.6 &201  &710 &1290 &9.96\\
\ \    &191.7 &7.5  &710 &1290 &9.97\\
&&&&&\\
\ \ CFC12 &293.7 &761.0 &800 &1270 &13.5\\
&&&&&\\
\ \ CFC113 &298.1 &760.0 &620 &5000 &14.6\\
&&&&&\\		
\ \ CFC114 &298.1 &760.0 &600 &5000 &17.4\\
&&&&&\\	
Hydrochlorofluorocarbons& & & & &\\
\ \ HCFC22  &294.8 &761.6 &730 &1380 &10.5\\
&&&&&\\	
\ \ HCFC141b&298.1 &760.0 &550 &6500 &8.57\\
&&&&&\\	
\ \ HCFC142b&298.1 &760.0 &600 &6500 &11.1\\
&&&&&\\			
Hydrofluorocarbons& & & & &\\
\ \ HFC23   &293.7 &762.4 &950 &1500 &12.3\\
&&&&&\\	
\ \ HFC125  &298.1 &760.0 &510 &6500 &17.5\\
&&&&&\\	
\ \ HFC134a &295.8 &760.8 &750 &1600 &13.2\\
&&&&&\\	
\ \ HFC143a &298.1 &760.0 &500 &6500 &14.1\\
&&&&&\\		
Fully Fluorinated Species& & & & &\\ 
\ \ SF$_6$   &189.2 &50.4   &780 &1100 &19.9\\
\ \    &298.1 &760.0  &560 &6500 &21.2\\
\ \    &323.1 &760.0  &560 &6500 &21.6\\
&&&&&\\
\ \ CF$_4$ &295.0 &760.5 &1190 &1336 &19.3\\
\hline
\end{tabular}
\end{center}
\caption{Hitran Absorption Cross Section Datasets Used in This Work.
\label{A}}
\end{table}

\noindent Here $N^{\{i\}}(z)$ is the density of a greenhouse gas molecule of type $i$ and $\sigma^{\{i\}}=\sigma^{\{i\}}(z,\nu)$ is its absorption cross section for radiation of frequency $\nu$ at the altitude $z$.  The cross section can depend strongly on altitude because temperature and pressure are functions of altitude.  Temperature controls the distribution of the molecules between translational, rotational and vibrational states.  Pressure, together with temperature, determines the width of the molecular resonance lines.  The halogenated gas absorption cross sections were obtained from the Hitran website \cite{HITRAN} and are listed in Table 1.  Halogens are very electronegative which implies that vibrations of halogenated molecules can have unusually large transition electric dipole moments, and therefore large absorption cross sections for thermal radiation.  The uncertainty of the Hitran cross sections is given to be about 8\% \cite{Buehler2022}.  Integrating over frequency gave the total cross sections listed in the final column of Table 1.  The total cross sections are believed to be independent of temperature \cite{Harrison2018}.  Table 1 shows the Hitran absorption cross sections for CFC11 and SF$_6$ vary less than 0.1\% and 2\% respectively, over a considerable range of temperatures and pressures.  This will be discussed more in Section 6. 

The cross section can be written as the sum of partial cross sections $\sigma_{ul}$, corresponding to each Bohr transition frequency $\nu_{ul}$,

\begin{equation}
	\sigma^{\{i\}}=\sum_{ul} \sigma_{ul}.
	\label{lbl16}
\end{equation}

\noindent The partial cross section, $\sigma_{ul}$, is the product of a  lineshape function, $G_{ul}=G_{ul}(\nu,\tau)$, and a
line strength, $S_{ul}=S_{ul}(T)$,
\begin{equation}
	\sigma_{ul}=G_{ul}S_{ul}.
	\label{lbl18}
\end{equation}

\noindent The absorption of the 5 main greenhouse gases, H$_2$O, CO$_2$, O$_3$, CH$_4$ and N$_2$O were obtained using the Hitran line strengths \cite{WvW1}.  

The optical depth from the surface to the top of the radiative atmosphere, the altitude $z_{\rm mp}$ of the mesopause, is

\begin{equation}
	\tau_{\infty}=\tau_{\rm mp}=\int_0^{z_{\rm mp}}dz'\kappa(z',\nu).
	\label{b11}
\end{equation}

\noindent As indicated by the notation (\ref{b11}), we have assumed that the optical depth $\tau_{\rm mp}$ at the mesopause altitude $z_{\rm mp}$ differs negligibly from the optical depth $\tau_{\infty}$ at infinite altitude since there is so little opacity of the atmosphere above the mesopause.

The Schwarzschild equation (\ref{nsa4}) can be solved to find the intensity \cite{WvW1}

\begin{eqnarray}
	\hbox{For $\varsigma>0$}:\quad\tilde I(\tau,\varsigma)&=& +\varsigma \int_0^{\tau}d\tau' e^{-\varsigma (\tau-\tau')}\tilde B(\tau')+e^{-\varsigma\tau}\tilde I(0,\varsigma)\label{vn48}\\
	\hbox{For $\varsigma<0$}:\quad\tilde I(\tau,\varsigma)&=& -\varsigma \int_{\tau}^{\tau_{\infty}}d\tau' e^{-\varsigma (\tau-\tau')}\tilde B(\tau')\label{vn50}
\end{eqnarray}

\noindent where $\varsigma= \sec \theta$.  

The upwards flux defined by 

\begin{equation}
	\tilde Z=\int_{4\pi}d\Omega\,\cos\theta\,\tilde I.
	\label{b5}
\end{equation}

\noindent can be rewritten after substituting (\ref{vn48}) and (\ref{vn50}) into (\ref{b5}) to give

\begin{eqnarray}
	{{\tilde Z}\over {2 \pi}}&=&\int_0^{\tau}d\tau'E_2(\tau-\tau')\tilde B(\tau') - \int_{\tau}^{\tau_{\infty}} d\tau' E_2(\tau'-\tau)\tilde B(\tau')+ \epsilon_s\tilde B_sE_3(\tau)  \nonumber\\
	&=&-\int_0^{\tau_{\infty}} d\tau' E_3(\vert \tau - \tau' \vert) {{\partial \tilde B(\tau')}\over{\partial \tau'}} +
	\tilde B(\tau_{\infty}) E_3(\tau_{\infty} - \tau).
	\label{vn14}
\end{eqnarray}

\noindent Here, $\tilde B_s=\tilde B(T_s)$, is the Planck intensity evaluated at the surface temperature $T_s$.  The exponential-integral functions, $E_n(\tau)$, that account for slant paths of radiation between different altitudes are defined for integers $n=1,2,3,\ldots$ by

\begin{equation}
	E_n(\tau)=\int_1^{\infty}d\varsigma\,\varsigma^{-n}\,e^{-\varsigma\tau}.
	\label{b40}
\end{equation}

\begin{figure}[t]   
	\includegraphics[height=100mm,width=1.0\columnwidth]{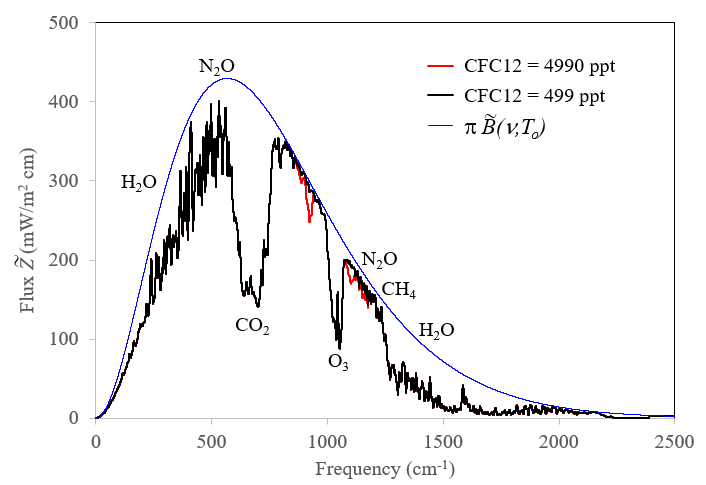} 
	\caption{Effects of different concentrations of CFC12 on the spectral flux at the mesopause altitude, $z_{\rm mp}$.  The smooth blue line is the spectral flux from a surface at the temperature of $T_s$ = 288.7 K without any greenhouse gases present.  Both the black and red lines show the flux for an atmosphere containing the 5 main greenhouse gases at their concentrations of Fig. \ref{GGNT} as well as CFC12.  The effect of CFC12 is only apparent if its concentration is 10 times its observed 2020 value, 499 ppt, as shown by the red curve.  
		\label{CFC12TOAFlux}}
\end{figure}

The spectral forcing, $\tilde F$, is defined as the difference between the spectral flux $\pi\tilde B_s$ through a transparent atmosphere from a surface with temperature $T_s$, and the spectral flux $\tilde Z$ for an atmosphere having greenhouse gases,

\begin{equation}
	\tilde F=\pi \tilde B_s-\tilde Z\\.
	\label{vn58}
\end{equation}

\noindent The spectral flux $\tilde Z$ at the mesopause altitude is shown in Fig. 4 for the case of the 5 main greenhouse gases as well as CFC12.

\begin{figure}[t]   
	\includegraphics[height=100mm,width=1.0\columnwidth]{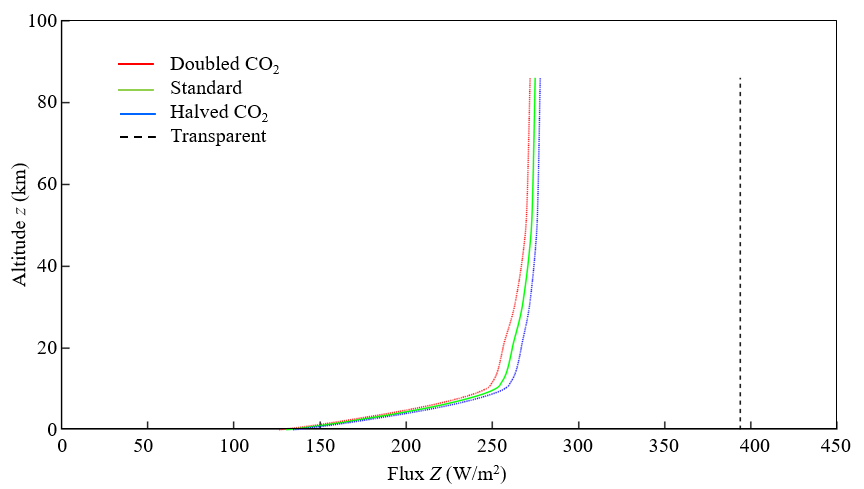} 
	\caption{Altitude dependence of frequency integrated flux $Z$ of (\ref{b58}).  The flux for three concentrations of CO$_2$ are shown, the standard concentration of Fig. \ref{GGNT}, twice and half that value. The other greenhouse gases have the standard concentrations of Fig. \ref{GGNT}. The vertical dashed line is the flux $\sigma_{\rm SB}T_s^4=394$ W m$^{-2}$ for a transparent atmosphere with a surface temperature $T_s = 288.7$ K.
		\label{ZzCM}}
\end{figure}

The frequency integrals of the flux (\ref{b5}) and the forcing (\ref{vn58}) are

\begin{eqnarray}
	Z&=&\int_0^{\infty}d\nu\,\tilde Z,\label{b58}\\
	F&=&\int_0^{\infty}d\nu\,\tilde F=\sigma_{\rm SB}T_s^4-Z,\label{b60}\\
\end{eqnarray}

\noindent where $\sigma_{SB}$ is the Stefan Boltzmann constant.  Fig. 5 shows the frequency integrated flux versus altitude for three different concentrations of CO$_2$ in addition to the four other main greenhouse gases.  A similar plot to Fig. 5 where one would show the result of increasing the concentration of any of the halogenated gases from its preindustrial level to its 2020 concentration results in an imperceptible change from the green curve.

\begin{sidewaystable}
	\begin{center}
		\begin{tabular}{|l|c|c|c|c|c|c|}
			\hline
			\ \ \ \ Molecule&$C^{\{i\}}$(2020) & $dC^{\{i\}}/dt$ &Altitude &Column Density $\hat N^{\{i\}}_{\rm sd}$ &\multicolumn{2}{c|}{Forcing (mW/m$^2$)} \\ 
			\ \ \ \ \ \ \ \ \ $i$ &(ppt)  &(ppt/yr) &Dependence &($10^{14}$ mol/cm$^2$)   &$z_{\rm tp}$   &$z_{\rm mp}$ \\
			[0.5ex]
			\hline\hline
			Chlorofluorocarbons&& & & & &\\
			\ \ CFC10 (CCl$_4$) &78  &-1.1 &No  &16.7 &20.6 &25.7\\
			\ \ CFC11           &226 &-1.9 &Yes &44.5 &84.6 &104\\
			\ \                 &    &     &No  &48.2 &91.3 &118\\
			\ \ CFC12           &499 &-3.8 &Yes &101  &219  &280\\
			\ \                 &    &     &No  &107  &227  &302\\
			
			\ \ CFC113          &69  &-0.64&No  &14.7 &32.3 &40.7\\
			\ \ CFC114          &16  &-0.02&No  &3.42 &7.64 &9.57\\
			
			Hydrochlorofluorocarbons& & & & & &\\
			\ \ HCFC22          &255 &2.2  &No  &54.4 &77.6 &102\\
			\ \ HCFC141b        &25  &0.10 &No  &5.34 &5.90 &7.40\\
			\ \ HCFC142b        &22  &-0.14&No  &4.70 &6.29 &8.39\\
			
			Hydrofluorocarbons& & & & & &\\
			\ \ HFC23           &38  &1.1  &No  &8.11 &10.4 &14.0\\
			\ \ HFC125          &32  &3.1  &No  &6.83 &10.9 &14.6\\
			\ \ HFC134a         &114 &6.1  &No  &2.43 &27.5 &36.7\\
			\ \ HFC143a         &25  &1.5  &No  &5.34 &6.07 &8.14\\
			
			Fully Fluorinated Species& & &  & & &\\ 
			\ \ SF$_6$          &10  &0.37 &Yes &1.97 &7.33 &9.18\\
			\ \                 &    &     &No  &2.13 &7.90 &10.5\\
			\ \ CF$_4$          &86 (34) &0.78 &No &18.4 &6.55 &8.75\\
			\hline
			& & & &Halogenated Gas Total &521 &667\\
			\hline
			Main Greenhouse Gases& & &  & & &\\ 
\ \ CO$_2$ &4.13 (2.80) $\times 10^8$ &$2.5 \times 10^6$ &      &$8.54 \times 10^7$ &2950 &1670\\
\ \ CH$_4$ &1.9 (0.8) $\times 10^6$ &$8 \times 10^3$ &Yes   &$3.73 \times 10^5$ &834  &761\\
\ \ N$_2$O &3.3 (2.7) $\times 10^5$ &$8 \times 10^2$ &Yes &$6.56 \times 10^4$ &277  &259\\			
\hline
\end{tabular}
\end{center}
\caption{Instantaneous Clear Sky Radiative Forcings at the altitude $z_{\rm tp} = 11$ km of the tropopause and $z_{\rm mp} = 86$ km of the mesopause. The total anthropogenic forcing was found relative to the preindustrial levels which was zero unless listed in brackets.  The observed rates of the concentration changes of the halogenated gases is given in Fig. 1 while the observed concentration changes for CH$_4$ and N$_2$O are discussed further in reference \cite{WvW3}.  The values for CFC11, CFC12 and SF$_6$ lacking altitude dependence were not considered when finding the halogenated gas total radiative forcings.  
\label{ab}}
\end{sidewaystable}

The radiative forcings of the various gases resulting from the increase in the respective concentration relative to their preindustrial levels are given in Table 2 at the tropopause and mesopause altitudes.  Table 2 as well as subsequent tables do not list water vapor, the most significant greenhouse gas.  As shown in Table 2 of reference \cite{WvW1}, the total radiative forcing of water vapor is about twice that of CO$_2$.  Clouds further increase the importance of H$_2$O for radiative transfer in Earth’s  atmosphere.  All halogenated gases except for CF$_4$ were not present in the atmosphere prior to the industrial revolution.  The effect of the altitude dependence on forcing was examined for CFC11, CFC12 and SF$_6$.  Taking a constant concentration at all altitudes increased their forcings by as much as 8\% at the tropopause and 14\% at the mesopause.  It is reasonable to assume that the radiative forcings of the other halogenated species are similarly affected.  

The effect due to all 14 halogenated species present in the atmosphere gives a forcing of 521 (667) mW/m$^2$ at the tropopause (mesopause) which is $\le$ 0.2\% of the total flux as shown in Fig. 5.  Nearly 60\% of these total forcings are due to CFC11 and CFC12 that have the highest halogenated gas concentrations.  The rate of change of each halogenated species was used to find the rate of change of the total halogenated gas radiative forcing to be 1.5 (2.2) mW/m$^2$/yr at the tropopause (mesopause).  This means the total halogenated gas radiative forcing will remain significantly less than that due to carbon dioxide and methane for the foreseeable future.

\section{Concentration Dependence of Forcing\label{cdp}}

The frequency integrated forcing, $F$, of (\ref{b60}) depends on the altitude $z$ and on the column densities of the five main greenhouse gases as well as the 14 halogenated species given in Table 2.
\begin{equation}
	F=F(z,\hat N^{\{1\}},\ldots,\hat N^{\{19\}}).
	\label{cdp8}
\end{equation}
We assume the temperature $T$ and densities $N^{\{i\}}$ have the same altitude profiles as shown in Fig. \ref{GGNT}.  An important special case of (\ref{cdp8}) is the forcing, $F_{\rm sd}$, when each greenhouse gas $i$ is present at its 2020 column density $\hat N^{\{i\}}_{\rm sd}$ of Table 2,

\begin{equation}
	F_{\rm sd}(z)=F(z,\hat N^{\{1\}}_{\rm sd},\ldots,\hat N^{\{19\}}_{\rm sd}).
	\label{cdp10}
\end{equation}
A second special case of (\ref{cdp8}) is the hypothetical, per molecule standard forcing, $F^{\{i\}}_{\rm sd}$, when the atmosphere contains only molecules of type $i$ at their standard column density, $\hat N^{\{i\}}=\hat N^{\{i\}}_{\rm sd}$, and the concentrations of the other greenhouse vanish, $\hat N^{\{j\}}=0$ if $j\ne i$,
\begin{equation}
	F^{\{i\}}_{\rm sd}(z)=F(z,0,\ldots,0,\hat N^{\{i\}}_{\rm sd}, 0,\ldots,0).
	\label{cdp12}
\end{equation}

We define the forcing power per added molecule as
\begin{equation}
	P^{\{i\}}(z,\hat N^{\{1\}},\ldots,\hat N^{\{n\}})=\frac{\partial F}{\partial \hat N^{\{i\}}}.
	\label{cdp14}
\end{equation}
The densities of greenhouse gases $j$ with $j\ne i$ are held constant in the
partial derivative of (\ref{cdp14}).  If the units of $F$ are taken to be W m$^{-2}$ and the units of $\hat N^{\{i\}}$ are taken to be molecules m$^{-2}$, then the units of $P^{\{i\}}$ will be W molecule$^{-1}$.

We define a finite forcing increment for the $i$th type of greenhouse molecule as
\begin{equation}
	\Delta F^{\{i\}}(z,f)=F(z,\hat N^{\{1\}}_{\rm sd},\ldots,\hat N^{\{i-1\}}_{\rm sd},f\hat N^{\{i\}}_{\rm sd}, \hat N^{\{i+1\}}_{\rm sd},\ldots,
	\hat N^{\{n\}}_{\rm sd})-F_{\rm sd}.
	\label{cdp18}
\end{equation}
\begin{table}
	\begin{center}
		\begin{tabular}{|l| c c | c c| c c |}
			\hline
			\ \ \ \ Molecule&\multicolumn{2}{c|}{$P^{\{i\}}_{\rm ot}(z)$}&\multicolumn{2}{c|}{$P^{\{i\}}_{\rm sd}(z,0)$}&\multicolumn{2}{c|}{$P^{\{i\}}_{\rm sd}(z,1)$}\\
			\ \ \ \ \ \ \ \ \ $i$& $z_{\rm tp}$ & $z_{\rm mp}$  & $z_{\rm tp}$ & $z_{\rm mp}$   & $z_{\rm tp}$  & $z_{\rm mp}$  \\ [0.5ex]
			\hline\hline
			&&&&&&\\
			Chlorofluorocarbons & & & & & &\\
			\ \ CFC10 (CCl$_4$) &1.62 &2.10 &1.25 &1.55 &1.24 &1.54 \\
			\ \ CFC11 &2.09 &2.64 &2.03 &2.41 &1.87 &2.31 \\
			\ \ CFC12 &2.36 &3.12 &2.29 &2.83 &2.12 &2.71 \\
			\ \ CFC113 &2.58 &3.50 &2.22 &2.82 &2.15 &2.74 \\
			\ \ CFC114 &2.77 &3.81 &2.31 &2.84 &2.25 &2.84 \\
			&&&&&&\\
			Hydrochlorofluorocarbons & & & & & &\\
			\ \ HCFC22 &1.69 &2.32 &1.42 &1.88 &1.42 &1.86 \\
			\ \ HCFC141b &1.52 &2.02 &1.11 &1.82 &1.10 &1.38 \\
			\ \ HCFC142b &1.69 &2.33 &1.39 &1.81 &1.33 &1.78 \\
			&&&&&&\\
			Hydrofluorocarbons & & & & & & \\
			\ \ HFC23 &1.60 &2.25 &1.29 &1.74 &1.28 &1.73 \\
			\ \ HFC125 &2.25 &3.13 &1.62 &2.14 &1.60 &2.13 \\
			\ \ HFC134a &1.66 &2.34 &1.13 &1.51 &1.13 &1.51 \\
			\ \ HFC143a & 1.70 &2.37 &1.14 &1.53 &1.14 &1.52 \\
			&&&&&&\\
			Fully Fluorinated Species & & & & & & \\
			\ \ SF$_6$ &4.28 &5.68 &3.75 &4.68 &3.71 &4.67 \\
			\ \ CF$_4$ &2.00 &2.85 &0.63 &0.81 &0.62 &0.80 \\
			&&&&&&\\
			Main Greenhouse Gases & & & & & & \\
			\ \ CO$_2$ \cite{WvW2} &0.276 &0.349 &0.22 &0.25 &$9.0 \times 10^{-5}$ &$4.9 \times 10^{-5}$ \\
			\ \ CH$_4$ \cite{WvW2}&0.052 &0.072 &0.021 &0.027 &$2.8 \times 10^{-3}$ &$2.6 \times 10^{-3}$ \\
			\ \ N$_2$O \cite{WvW2}&0.170 &0.227 &0.073 &0.091 &$2.1 \times 10^{-2}$ &$2.0 \times 10^{-2}$ \\
			\hline
		\end{tabular}
	\end{center}
	\caption{Forcing powers (\ref{cdp14}) per additional molecule in units of $10^{-21}$ W at the altitude $z_{\rm tp} = 11$ km of the tropopause and $z_{\rm mp} = 86$ km of the mesopause. Altitude dependences were taken into account as given in Fig. 2 and Fig. 3 while all other species had altitude independent concentrations. The surface temperature was $T_s=288.7$ K, and the altitude profile of temperature is given in  Fig. $\ref{GGNT}$.  $P^{\{i\}}_{\rm ot}(z)$ of (\ref{ot4}) is for the optically-thin limit.  $P^{\{i\}}_{\rm sd}(z,0)$ from (\ref{cdp20}) is for an atmosphere that previously had no molecules of type $i$ (so $\hat N^{\{i\}}=0$) but all other greenhouse molecules had standard concentrations.  $P^{\{i\}}_{\rm sd}(z,1)$ from (\ref{cdp20}) is for a single molecule of type $i$  added to an atmosphere that previously had standard  densities for all 5 main greenhouse gases.\label{dPr}}
\end{table}

Differentiating (\ref{cdp18}) with respect to $f$ we find
\begin{equation}
	\frac{\partial \Delta F^{\{i\}}}{\partial f}(z,f)=\hat N^{\{i\}}_{\rm sd}P^{\{i\}}_{\rm sd}(z,f),
	\label{cdp20}
\end{equation}
where $P^{\{i\}}_{\rm sd}(z,f)$ is the forcing power per additional molecule of type $i$ when these molecules have the column density $\hat N^{\{i\}}=f\hat N^{\{i\}}_{\rm sd}$ and all other types of greenhouse molecules have their standard column densities.

We now consider the optically thin limit, where the concentrations of greenhouse gases are sufficiently low that the optical depths $\tau$ of (\ref{b10}) will be small, $\tau\ll 1$, for all frequencies $\nu$ and at all altitudes $z$.  The frequency integral of the spectral forcing (\ref{b60}) at altitude $z$ can then be written as

\begin{eqnarray}
	F_{\rm ot}(z)&=&\sum_i\hat N^{\{i\}} P^{\{i\}}_{\rm ot}(z)
	\label{ot2}
\end{eqnarray}

\noindent where the forcing power per molecule of type $i$ is

\begin{equation}
	P^{\{i\}}_{\rm ot}(z)=\frac{1}{2}\int_0^z dz'\frac{N^{\{i\}'}}{\hat N^{\{i\}}}\left[\Pi^{\{i\}}(T',T_s)-\Pi^{\{i\}}(T',T')\right]
	+\frac{1}{2}\int_z^{\infty} dz'\frac{N^{\{i\}'}}{\hat N^{\{i\}}}\Pi^{\{i\}}(T',T').
	\label{ot4}
\end{equation}

\noindent Here $N^{\{i\}'} = N^{\{i\}}(z')$, and $T'=T(z')$. 
The mean power absorbed by a greenhouse gas molecule of temperature $T$ from thermal equilibrium radiation of temperature $T'$ is

\begin{equation}
	\Pi^{\{i\}}(T,T')= 4\pi\sum_{ul} S_{ul}^{\{i\}}(T)  \tilde B(\nu_{ul},T').
	\label{ot6}
\end{equation}

\noindent For $z'>z$ we see from (\ref{ot4}) that the $dz'\, N'$ molecules in the altitude interval $z'$ to $z'+dz'$ each emit the power $\Pi(T',T')$, of which half goes to outer space and half goes down through the reference plane at altitude $z$, diminishing the net flux through the reference plane by $\Pi(T',T')/2$.  Molecules above the reference plane can only cause positive forcing unlike molecules below the reference plane which can cause either positive or negative forcing.

An important check of our work was to use (\ref{ot4}) to calculate the forcing power per molecule for the optically thin limit.  The results displayed in column 2 of Table 3 agreed with that found by dividing the radiative forcing in the limit of zero gas concentration, by the column density.

The forcing powers per molecule given in Table 3 are all on the order of $10^{-21}$ W for the halogenated gases.  The values are largest for the optically thin case and slightly lower for $P^{\{i\}}_{\rm sd}(z,0)$ and $P^{\{i\}}_{\rm sd}(z,1)$ because of saturation.  The effect of satuation is much larger for the 5 main greenhouse gas values as explained in reference \cite{WvW2} because their number densities are up to $10^{7}$ times greater than those of the halogenated gases.  

The forcing power can be used to estimate the contribution of each gas to global warming as follows.  First, one assumes that the temperature increase $\Delta T^{\{i\}}$ due to a change in the concentration of gas $i$ is proportional to its radiative forcing change,  

\begin{equation}	
\Delta T^{\{i\}} = c \ \Delta F^{\{i\}}
	\label{ff26b}
\end{equation}

\begin{table}
	\begin{center}
		\begin{tabular}{|l|c|}
			\hline
			&\\	
			\ \ \ \ Molecule &$\Delta F^{\{i\}}/{\Delta F_{\rm tot}}$ \\ 
			&(\%)\\
			\hline\hline
			&\\
			Chlorofluorocarbons &\\
			\ \ CFC10 (CCl$_4$) &-0.51\\
			\ \ CFC11           &-1.32\\
			\ \ CFC12           &-2.99\\	
			\ \ CFC113          &-0.51\\
			\ \ CFC114          &-0.02\\
			&\\			
			Hydrochlorofluorocarbons&\\
			\ \ HCFC22          &1.16\\
			\ \ HCFC141b        &0.04\\
			\ \ HCFC142b        &-0.07\\
			&\\			
			Hydrofluorocarbons&\\
			\ \ HFC23           &0.52\\
			\ \ HFC125          &1.84\\
			\ \ HFC134a         &2.56\\
			\ \ HFC143a         &0.63\\
			&\\			
			Fully Fluorinated Species&\\ 
			\ \ SF$_6$          &0.51\\
			\ \ CF$_4$          &0.18\\	
			&\\
			Main Greenhouse Gases&\\ 
			\ \ CO$_2$          &83.4\\
			\ \ CH$_4$          &8.3\\
			\ \ N$_2$O          &6.2\\		
			\hline
			&\\
			Total &100.0\\
			\hline
		\end{tabular}
	\end{center}
	\caption{Relative Contribution of Greenhouse Gases to Total Radiative Forcing Change.  $\Delta F_{\rm {Tot}}$ represents the total radiative forcing change caused by changes in the concentrations of all greenhouse gas molecules.  The relative forcing was found using the observed 2020 concentration change rate of each gas $i$ as discussed in the text.  A negative relative forcing change occurs if the halogenated gas atmospheric concentration is decreasing as a result of the Montreal Protocol.}
	\label{Tab6}
\end{table}

\noindent where $c$ is a constant of proportionality whose estimation is difficult because it includes feedback effects.  The radiative forcing change at the tropopause due to a concentration change occurring over time $\Delta t$ is given by 

\begin{equation}	
\Delta F^{\{i\}} = {P^{\{i\}}_{\rm sd}(z_{\rm tr},1)} {{dC^{\{i\}}}\over {dt}} \Delta t.
\label{ff26c}
\end{equation}

\noindent Here, we have evaluated the forcing power per molecule at the tropopause which is given in Table 3.  The total radiative forcing change due to changes in all $i$ greenhouse gases over time $\Delta t$ is given by

\begin{equation}	
	\Delta F_{\rm Tot}	 = \sum_i \Delta F^{\{i\}}.
	\label{ff26e}
\end{equation}

\noindent Table 4 gives the relative contributions of the various gases to the total radiative forcing change.  This was found evaluating (\ref{ff26c}) and (\ref{ff26e}) using the 2020 observed concentration change rates given in Table 2.  The halogenated gas total is responsible for 2\% of the total radiative forcing change. 

The total temperature change due to changes in all $i$ greenhouse gas concentrations over time $\Delta t$ is given by

\begin{equation}	
	\Delta T_{\rm Tot}	 = \sum_i \Delta T^{\{i\}}.
	\label{ff26d}
\end{equation}

\noindent Equations (\ref{ff26b}), (\ref{ff26e}) and (\ref{ff26d}) imply

\begin{equation}	
	{{\Delta T^{\{i\}}} \over {\Delta T_{\rm Tot}}}	 = {{\Delta F^{\{i\}}}\over {\Delta F_{\rm Tot}}}.
	\label{ff26f}
\end{equation}

\noindent Surface temperatures have increased by about 1.1 K since 1850 according to the IPCC \cite{IPCC2021}.  More recent satellite measurements since 1979, show a slightly higher warming rate of about 0.1 C per decade \cite{Spencer}.  Since the observed temperature increase is comparable to previous temperature increases when there were negligible changes of greenhouse gas concentrations, there is no way to  know how much is due to currently increasing greenhouse gases. But if one makes the extreme assumption that changes in greenhouse gas concentrations have driven all of this observed warming, one finds the changes in the halogenated gas concentrations are responsible for a warming of 0.002 C per decade. 

\section{Global Warming Potential}

A comparison of the greenhouse warming contribution of a gas $i$ relative to carbon dioxide is provided by the global warming potential which is defined as \cite{WvW3}

\begin{equation}	
	\hbox{GWP}^{\{i\}}(T) = \frac{\langle \hbox{RF}^{\{i\}}\rangle}{\langle \hbox{RF}^{\{{\rm CO}_2\}}\rangle}.
	\label{gwp2}
\end{equation}

\noindent The time-integrated radiative forcing at the tropopause per unit mass over an observation time $T$ is

\begin{equation}	
	\langle \hbox{RF}^{\{i\}}\rangle=\frac{P^{\{i\}}\theta^{\{i\}}(T)}{m^{\{i\}}}.
	\label{gwp4}
\end{equation}

\noindent The units of $\langle \hbox{RF}^{\{i\}}\rangle$ are joules per square meter per unit mass.  It is the amount of ``forcing heat" acquired by a square-meter column of the atmosphere during time $T$ after the pulse emission of one unit mass of greenhouse gas of type $i$.
The forcing time $\theta^{\{i\}}$ is
\begin{equation}	
	\theta^{\{i\}}(T)=\int_0^T dt' f^{\{i\}}(t').
	\label{gwp6}
\end{equation}
$f^{\{i\}}(t')$ is the fraction of the excess greenhouse gas molecules remaining in the atmosphere at a time $t'$ after the ``pulse emission.''  Excess greenhouse molecules can remain in the atmosphere long after the greenhouse molecules of the original pulse have exchanged with the land and oceans because they are replaced with equivalent molecules that continue to provide radiative forcing. An example is the relatively short time needed for  the exchange of $^{14}$CO$_2$ molecules from atmospheric nuclear weapon tests with CO$_2$ molecules of the ocean. It takes a much longer time for excess CO$_2$ concentrations to decay away. The excess atmospheric concentration of molecules, caused by the emitted pulse of the species $i$, can be removed from the atmosphere by various mechanisms. For example, CO$_2$ is absorbed by the biosphere and oceans.  CH$_4$ is oxidized by OH radicals and other atmospheric gases. N$_2$O and the various halogenated gases are photodissociated by solar ultraviolet radiation in the stratosphere, etc.  

It is convenient to describe the excess fraction $f^{\{i\}}(t)=f(t)$ with a sum of $n+1$ exponentially decaying components, of amplitudes $a_j$ and time constants $\tau_j$\,\cite{Harvey, Joos}
\begin{equation}	
	f(T) =\sum_{j=0}^n a_j e^{-T/\tau_j},\quad\hbox{where}\quad \sum_{j=0}^n a_j = 1.
	\label{gwp8}
\end{equation}
Substituting (\ref{gwp8}) into (\ref{gwp6}) we find that the forcing time, $\theta^{\{i\}}=\theta$, is
\begin{equation}	
	\theta(T)=\sum_{j=0}^n a_j\tau_j\left(1-e^{-T/\tau_j}\right).
	\label{gwp10}
\end{equation}
For the limiting case of infinitely long time constants, $\tau_j\to\infty$, one should make the replacement $\tau_j\left(1-e^{-T/\tau_j}\right)\to T$ in (\ref{gwp10}). From inspection of (\ref{gwp10}) one can conclude that
the forcing time $\theta^{\{i\}}$ is always less than or equal to the observation time, $\theta^{\{i\}}(T)\le T$.
Using (\ref{gwp10}) in (\ref{gwp4}) we find that the global warming potential is
\begin{equation}	
	\hbox{GWP}^{\{i\}}(T)= \left(\frac{P^{\{i\}}}{P^{\{{\rm CO}_2\}}}\right)\left(\frac{m^{\{{\rm CO}_2\}}}
	{m^{\{i\}}}\right)\left(\frac{\theta^{\{i\}}(T)}{\theta^{\{{\rm CO}_2\}}(T)}\right).
	\label{gwp12}
\end{equation}
\begin{sidewaystable}
	\begin{center}
		\begin{tabular}{|l|c|c|c|c|c|c|c|}
			\hline
			
			\ \ \ \ Molecule &Mass &Lifetime\cite{IPCC2021Sup} &\multicolumn{3}{c|}{This Work} &\multicolumn{2}{c|}{IPCC 2021 \cite{IPCC2021Sup}} \\ 
			\ \ \ \ \ \ \ \ \ $i$&$m^{\{i\}}$(amu)  &$\tau_i$(yr) &GWP(0) &GWP(20)   &GWP(100)   &GWP(20) &GWP(100)\\
			
			\hline\hline
			&&&&&&&\\
			Chlorofluorocarbons && & & & & &\\
			\ \ CFC10 (CCl$_4$) &151.1    &32  &4296 &4179  &2494 &3810 &2200\\
			\ \ CFC11           &137.4    &52  &6654 &7225  &5605 &7430 &5560\\
			\ \ CFC12           &120.9    &102 &8573 &10182 &10367 &11400  &11200\\	
			\ \ CFC113          &187.4    &93  &5609 &6601  &6520 &6860 &6520\\
			\ \ CFC114          &170.9    &89  &6437 &7540  &7336 &8260 &9430\\
			&&&&&&&\\			
			Hydrochlorofluorocarbons& & & & & & &\\
			\ \ HCFC22          &86.5    &11.9 &8026  &5083  &1812 &5690 &1960\\
			\ \ HCFC141b        &116.9   &9.4  &4600  &2491  &820  &2710 &860\\
			\ \ HCFC142b        &100.5   &18   &6470  &5109  &2201 &5510 &2300\\
			&&&&&&&\\			
			Hydrofluorocarbons& & & & & & &\\
			\ \ HFC23           &70    &228  &8940  &11195  &13732 &12400 &14600\\
			\ \ HFC125          &120   &30   &6519  &6223   &3578  &6740  &3740\\
			\ \ HFC134a         &102   &14   &5416  &3770   &1438  &4140 &1530\\
			\ \ HFC143a         &84    &51   &6635  &7179   &5517  &7840 &5810\\
			&&&&&&&\\			
			Fully Fluorinated Species& & & & & & &\\ 
			\ \ SF$_6$          &146   &3200 &12423 &16198 &23207 &18300 &25200\\
			\ \ CF$_4$          &88    &50000 &3444 &4504  &6529  &5300 &7380\\	
			&&&&&&&\\
			Main Greenhouse Gases& & & & & & &\\ 
			\ \ CO$_2$           &44  &See Caption &1 &1 &1 &1 &1\\
			\ \ CH$_4$ \cite{WvW3}         &16  &11.8  &85.5 &53.9 &19.2 &81.2 &27.9\\
			\ \ N$_2$O \cite{WvW3}         &44  &109   &233  &279  &290  &273  &273\\		
			\hline
		\end{tabular}
	\end{center}
	\caption{Comparison of Global Warming Potentials calculated for this work using equation (\ref{gwp12}) and the tropopause powers given in Table 3 with IPCC 2021 values.  There are multiple pathways for CO$_2$ to be removed from the atmosphere as is discussed in the text.}
	
	\label{ab2}
\end{sidewaystable}

Unlike the forcing powers $P^{\{i\}}$ per greenhouse molecule of type $i$, which can be accurately calculated, the time dependence of the fractions $f^{\{i\}}(T)$ is not well known. So, the global warming potentials GWP$^{\{i\}}$ are of limited quantitative value and do not contribute in a clear way to estimating the warming caused by a change in a greenhouse gas concentration. To add further uncertainty, ``indirect effects" are sometimes included in the GWP$^{\{i\}}$, for example, the effects of the CO$_2$ and H$_2$O molecules which result from the oxidation of CH$_4$.  For this paper only the direct forcing of the greenhouse gases was considered. 

For CO$_2$, Harvey\,\cite{Harvey} uses a five-exponential form of (\ref{gwp8}) with the parameters
\begin{equation}
	\left[\begin{array}{c}a_0\\ a_1\\
		a_2\\ a_3 \\ a_4\end{array}\right]
	=\left[\begin{array}{c}.131\\ .201\\ .321\\ .249 \\.098\end{array}\right]\quad \hbox{and}\quad
	\left[\begin{array}{c}\tau_0\\ \tau_1\\
		\tau_2\\ \tau_3 \\ \tau_4\end{array}\right]
	=\left[\begin{array}{c}\infty\\ 362.9\\ 73.6\\ 17.3 \\ 1.9\end{array}\right]\hbox{ yr}.
	\label{gwp16}
\end{equation}
Alternate parameterizations of $f^{\{{{\rm CO}_2}\}}$, for example, those of Joos {\it et al.}\,\cite{Joos}, give GWPs that do not differ more than the uncertainties of the estimates. 

Table 5 shows the global warming potentials calculated with (\ref{gwp12}) using the lifetimes of the various greenhouse gases given in the most recent IPCC report \cite{IPCC2021Sup} and equation (\ref{gwp16}).  Our GWP values are consistent with those of the 2021 IPCC report.  The somewhat larger values of GWP$^{\{\rm CH_4\}}$ may be due to indirect effects included in the IPCC calculations.

\section{Summary}

This work found the instantaneous clear sky radiative forcings of 14 halogenated molecules that have previously been found to exert the largest greenhouse warming.  CFC11 and CFC12 comprise about 60\% of the total forcing generated by the 14 halogenated molecules and their concentrations have been decreasing in recent years at 2\% and 4\% per year, respectively.  The concentrations of 6 of the 14 halogenated gases studied are decreasing as a result of the Montreal Protocol.

A number of factors affect the accuracy of the results.  First, the calculations used the absorption cross sections given by the Hitran database which have an uncertainty of several percent.  Second, the Hitran data does not show a temperature dependence of the cross sections.  This contrasts with our previous work \cite{WvW2} that examined the effect of temperature on the frequency integrated cross section which equals the sum of the line strengths.  Negligible temperature effect was found for the 5 main greenhouse gases as well as CF$_4$ but the absorption cross section for SF$_6$ increased by a factor of 2 as temperature decreased from 300 to 200 K.  Molecules made of many atoms, like SF$_6$, have many low-frequency, infrared-inactive modes which can be significantly excited at higher temperatures \cite{WvW2}.  Then the molecule spends more time in nonradiating modes and emits less average power at the frequencies of infrared active modes than a simpler molecule would.  The temperature increase causes a corresponding decrease in the absorption cross section of the infrared active mode.  The absorption cross sections of other complicated halogenated moecules can be expected to have analogous temperature dependences.  Finally, the altitude dependence of the halogenated gas concentrations is poorly known.  For the cases of CFC11, CFC12 and SF$_6$, the radiative forcing increases by up to 14\% if the altitude concentration dependence is ignored.  

Our calculations were done for the case of a clear sky.  In reality, about 70\% of the Earth's surface is covered by clouds \cite{ISCCP}.  A cloud layer will particularly perturb the so called spectral window shown between the CO$_2$ and O$_3$ absorptions in Fig. 4, where the halogenated gas absorptions most affect the radiative forcing.  It is therefore reasonable that incorporating clouds, will reduce the halogenated gas radiative forcing.  Some studies have found clouds reduce the halogenated gas forcing by 25\% \cite{Buehler2022}.

This study computed the global warming potentials which are comparable to those found by the IPCC2021 as shown in Table 5.  It also evaluated the forcing power per molecule defined as the derivative of the radiative forcing of a gas with respect to its column density.  The advantage of the per-molecule forcing power (\ref{cdp14}) is that unlike the GWP it enables one to quickly estimate the relative warming caused by a concentration change of a greenhouse gas.  The forcing power strongly depends on the gas concentration.  At very low concentrations, all gases have an optically thin power on the order of $10^{-21}$ W per molecule.  Table 3 shows the forcing power per molecule of the halogenated gases decreases slightly at higher gas concentrations.  However, for the 5 main greenhouse gases which have much higher gas concentrations, the forcing power per molecule decreases by orders of magnitude.  The relative radiative forcing change caused by a change in the concentration of each gas is given in Table 4.  Most of the total radiative forcing change is caused by increases in CO$_2$ (83\%), CH$_4$ (8\%) and N$_2$O (6\%).  The radiative forcing change due to the 14 halogenated molecules is only 2\% of the total.  Assuming the temperature change due to a changing gas concentration is proportional to its radiative forcing change, the halogenated gases are responsible for a surface warming of 0.002 C per decade compared to the total observed warming of about 0.1 C per decade.  

\section*{Acknowledgements}
The Canadian Natural Science and Engineering Research Council provided financial support of one of us.

\end{document}